\documentclass[sigconf,nonacm]{acmart}

\usepackage{tcolorbox}
\tcbuselibrary{breakable, skins}
\usepackage{listings}
\usepackage{tikz}

\usetikzlibrary{positioning, arrows.meta, shapes.geometric}

\AtBeginDocument{%
  }

\setcopyright{acmlicensed}
\copyrightyear{2018}
\acmYear{2018}
\acmDOI{XXXXXXX.XXXXXXX}
\acmConference[Conference acronym 'XX]{Make sure to enter the correct
  conference title from your rights confirmation email}{June 03--05,
  2018}{Woodstock, NY}
\acmISBN{978-1-4503-XXXX-X/2018/06}

\begin{document}

\title{Simulating the Resident: Generating Executable Smart Home Schedules via LLM Personas}

\author{Victor Jüttner}
\email{victor.juettner@uni-leipzig.de}
\orcid{0009-0005-7290-1996}
\author{Erik Buchmann}
\email{erik.buchmann@uni-leipzig.de}
\orcid{0009-0009-5874-4313}
\affiliation{%
  \institution{ScaDS.AI Dresden/Leipzig, Leipzig University}
  \city{Leipzig}
  \country{Germany}
}

\author{Xenia Wagner}
\email{xenia.wagner@rohde-schwarz.com}
\author{Christoph Jahn}
\email{christoph.jahn@rohde-schwarz.com}
\affiliation{%
  \institution{ipoque GmbH, a Rohde \& Schwarz company}
  \city{Leipzig}
  \country{Germany}
}


\renewcommand{\shortauthors}{J{\"u}ttner et al.}


\begin{abstract}
Smart homes have emerged as an important domain for HCI research, including work on usable security and privacy. Ideally, studies in these areas draw on datasets collected in real homes with real residents, capturing authentic device interactions, network traffic, and daily routines. However, creating such datasets is slow, expensive, and raises significant privacy concerns, as it requires long-term observation of people in their most private spaces. We propose using LLMs to generate diverse resident personas that interact with a simulated smart home, producing behaviorally grounded interaction schedules that can be executed on physical testbeds. We present (1) a design framework configuring simulated households across five socio-technical dimensions, (2) a multi-stage LLM pipeline that produces structured, executable device interaction schedules, and (3) a proof of concept demonstrating feasibility. As a work in progress, we aim to support scalable, privacy-conscious smart-home experimentation without relying on intrusive real-world data collection.
\end{abstract}




\maketitle

\begingroup 

\renewcommand\thefootnote{} 

\footnotetext{\scriptsize AI-HCD 2026: 1st Symposium on Artificial Intelligence throughout the Human-Centered Design Process, HTWD -- University of Applied Sciences, 01069 Dresden, Germany. \\ © 2026 Copyright held by the owner/author(s), DOI: \url{https://doi.org/10.18420/AIHCD2026_025}. \\ Except as otherwise noted, this paper is licenced under the Creative Commons Attribution 4.0 International Licence. To view a copy of this licence, visit \url{http://creativecommons.org/licenses/by/4.0}. } 

\endgroup 

\section{Introduction}

Smart homes are becoming an integral part of everyday domestic life, with the number of connected households growing rapidly worldwide~\cite{fortune_smart_home_2026}. 
Smart devices such as lights, thermostats, and speakers generate continuous traces of how residents live, move, and automate their daily routines~\cite{cook2013casas, zeng2017end}, making smart homes an increasingly important setting for HCI research. Behavioral traces from smart devices are relevant for designing usable, context-aware systems, but also carry significant implications for privacy and security: Observable device traffic can reveal fine-grained information about residents' habits and daily behavior~\cite{apthorpe2019keeping, trimananda2020packet, acar2020peekaboo}. 

However, studying realistic smart-home behavior constitutes a methodological challenge. Researchers need data about how people actually interact with devices in everyday life in order to design and evaluate secure and privacy-aware systems, automation logic, and user experiences. Yet collecting such data in real households often requires long-term and invasive observation, creating a tension between the need for ecologically valid data and protecting residents' privacy.
 
Real smart-home datasets~\cite{cook2013casas, alemdar2013aras, neto2023ciciot2023} are expensive to collect, limited to a small number of homes, and do not easily generalize across diverse household compositions, device ecosystems, or usage patterns. One way to bypass these limitations is to simulate household behavior synthetically. Large language models (LLMs) offer a promising approach: They have recently demonstrated strong potential for generating believable personas, structured narratives, and context-sensitive daily routines~\cite{park2023generative}.

We therefore propose using LLMs to simulate the residents themselves. By modeling diverse personas and their daily routines, it becomes possible to construct arbitrarily varied smart-home scenarios, different device ecosystems, household compositions, and dynamics between residents. Crucially, these simulated routines are not limited to text-based behavior logs: They can be translated into concrete interaction schedules and executed on physical smart-home testbeds, for example via home automation platforms like Home Assistant~\cite{homeassistant2026} or smartphone-based automation systems such as scrcpy by Genymobile~\cite{genymobile-scrcpy} or the Test \& Training Data Automation System (TTDAS) by Rohde \& Schwarz. This allows to capture authentic network traffic and device state changes without subjecting real households to intrusive observation.
Based on this idea, we formulate our central research question: 
\textbf{How can an LLM pipeline transform household context and resident personas into executable smart-home interaction schedules for physical testbeds?}

We approach this question in three parts:

\textbf{(1) A design framework} that configures simulated households with five socio-technical dimensions: Occupational routines, simulation timeframe, household dynamics, device ecosystem, and environmental context. 

\textbf{(2) A multi-stage LLM pipeline} based on the generative agent architecture introduced by Park et al.~\cite{park2023generative}, adapted to generate resident personas, narrative day-plans, and structured, executable device interaction schedules for physical smart-home testbeds. 

\textbf{(3) A proof-of-concept demonstration} showing that the pipeline produces temporally coherent, persona-consistent, and schema-compliant interaction traces.

As a work in progress, this paper focuses on framework design and this initial demonstration rather than a fully validated method, with the broader goal of supporting scalable, privacy-conscious smart-home experimentation without relying on intrusive real-world data collection.

\section{Related Work}


To bypass the cost and privacy constraints of real deployments, researchers have explored synthetic data generation. Early approaches such as SynSys~\cite{dahmen2019synsys} and SynD~\cite{klemenjak2020synd} use statistical models and Hidden Markov Models to generate artificial sensor and energy traces useful for data augmentation, but neither incorporates a semantic model of human intent. Park et al.\ introduced generative agents~\cite{park2023generative}: LLM-driven personas that produce believable daily routines grounded in social and environmental context, demonstrating that LLMs can simulate human behavior at a level of nuance that statistical models cannot reach. Building on this, Leng et al.\ propose AgentSense~\cite{leng2025agentsense}, which combines LLM-generated routines with 3D virtual environments to synthesize ambient sensor data --- but remains confined to virtual settings with no connection to real hardware. Most closely related to our work, Xu et al.\ propose IoTGen~\cite{xu2025iotgen}, which uses LLMs to transform existing real-world IoT behavior sequences into new text-based datasets adapted to different household configurations and seasons.

While IoTGen demonstrates that LLMs can meaningfully enrich behavioral data, it relies on existing real-world sequences as input and produces text-based output that cannot directly drive physical devices. Our work takes a different path: We simulate residents from scratch using only a configured persona and household context, and translate the resulting routines into executable interaction schedules for physical smart-home testbeds --- enabling researchers to capture authentic device responses and network traces without any prior real-world data collection.

\section{Simulation Framework}

Our approach consists of two integrated components: We define a design framework that allows to configure five socio-technical household dimensions and a multi-stage LLM pipeline that translates these dimensions into executable device interactions.

\subsection{Design Framework for Households}

To create realistic and configurable smart-home simulations, we derive five socio-technical dimensions from literature, allowing to shape simulated device activity. The dimensions are informed by prior work on household routines, multi-user coordination, and smart-home interaction patterns.

\textbf{Occupational routines} capture how residents' work patterns structure daily life and, in turn, influence when they are at home and likely to use smart-home devices~\cite{cook2013casas}. Example occupations with distinct temporal activity patterns are a \textit{9-to-5 office worker}, a \textit{work-from-home professional}, and a \textit{shift worker}.

\textbf{Simulation timeframe} defines the temporal scope of the generated data, shaping whether the simulation captures standard daily routines, weekday-to-weekend variations, or long-term habit changes~\cite{dahmen2019synsys, garcia2020iot23}. Example scopes include a \textit{single-day snapshot}, a \textit{full week}, or a \textit{longitudinal} simulation run.

\textbf{Household dynamics} capture the social structure of the home and characterize how residents coordinate, overlap, or interfere with one another's device usage~\cite{geeng2019whos, zeng2019multiuser}. Example configurations are a \textit{single resident}, \textit{cooperative cohabitants} with aligned routines, and \textit{conflicting schedules} where residents must negotiate shared spaces.

\textbf{Device ecosystem and interaction style} reflect the installed hardware and technological literacy of the household, determining the complexity and automation level of the generated interactions~\cite{chi2023detecting, yao2019bystanders}. Example setups are a \textit{basic setup} reliant on manual smartphone toggles, or a \textit{power-user setup} driven by complex background automations.

\textbf{Environmental context} situates routines within a specific physical setting and climate, influencing spatial behaviors and environmentally triggered device usage like heating or lighting. Example contexts include a \textit{compact urban apartment}, a spacious \textit{suburban family home}, and specific \textit{seasonal conditions} such as a cold winter morning.

These dimensions are deliberately simple and extensible. They provide a compact starting point for generating varied but comparable simulation scenarios across different research settings.

\subsection{Simulation Pipeline}

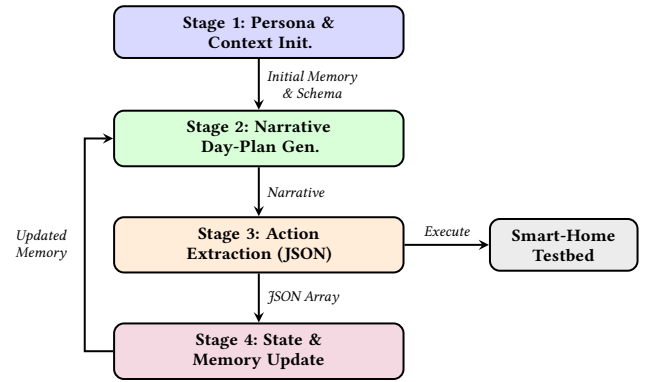
\begin{figure}[htbp]
\centering
\resizebox{\columnwidth}{!}{%
\begin{tikzpicture}[
    stage/.style={rectangle, rounded corners, draw=black, thick, text width=3.8cm, align=center, font=\footnotesize\bfseries, inner sep=4pt},
    data/.style={align=center, font=\scriptsize\itshape},
    arrow/.style={thick, ->, >=stealth}
]

\node (s1) at (0, 0) [stage, fill=blue!15] {Stage 1: Persona \&\\Context Init.};
\node (s2) at (0, -1.5) [stage, fill=green!15] {Stage 2: Narrative\\Day-Plan Gen.};
\node (s3) at (0, -3.0) [stage, fill=orange!15] {Stage 3: Action\\Extraction (JSON)};
\node (s4) at (0, -4.5) [stage, fill=purple!15] {Stage 4: State \&\\Memory Update};

\node (testbed) at (4.3, -3.0) [stage, fill=gray!15, text width=1.8cm] {Smart-Home\\Testbed};

\draw [arrow] (s1) -- node[data, right] {Initial Memory\\\& Schema} (s2);
\draw [arrow] (s2) -- node[data, right] {Narrative} (s3);
\draw [arrow] (s3) -- node[data, right] {JSON Array} (s4);

\draw [arrow] (s3) -- node[data, above] {Execute} (testbed);

\draw [arrow] (s4.west) -- ++(-0.4,0) |- node[data, near start, left, text width=1.0cm] {Updated\\Memory} (s2.west);

\end{tikzpicture}%
}
\vspace{-2mm} 
\caption{Overview of the four-stage simulation pipeline.}
\label{fig:pipeline-flowchart}
\end{figure}

Our pipeline, shown in~\autoref{fig:pipeline-flowchart}, adapts the memory--planning pattern from generative agents~\cite{park2023generative} to a device-centric smart-home setting. The four stages form a loop that translates a chosen configuration into execution-ready smart-home interaction traces.

\paragraph{Stage 1: Persona and Context Initialization}

In the first stage, the pipeline initializes the simulation using the chosen household configuration. As shown in~\autoref{fig:prompt-stage-1}, the LLM is provided with the residents' occupational routines and household dynamics, a global context defining the environmental setting and simulation timeframe, and a pre-defined JSON schema representing the available devices and their capabilites. Based on these inputs, the model generates a concise ``persona memory card'' for each resident. This profile summarizes their daily schedule and smart-home habits while explicitly mapping their preferences to exact device names and specific parameter values from the schema.

\begin{center}
\centering
\begin{tcolorbox}[breakable, enhanced, sharp corners, colback=white, colframe=blue!50, boxrule=1.0pt, left=1mm, right=1mm, top=1mm, bottom=1mm, nobeforeafter]
\small
You are simulating a smart home occupied by \texttt{[N]} residents. The residents are:\ \texttt{[resident descriptions including occupational routines and household dynamics].}

\medskip

Global Context: \texttt{[description of dwelling, environmental, and temporal context]}.

\medskip

The home contains the following devices and allowed actions only:\
\texttt{[device schema JSON with separated actions and value formats]}

\medskip

For each resident, produce a concise persona memory card summarizing their daily schedule, smart-home habits, and relevant preferences. State their preferences using the exact device names and specific parameter values (e.g., "prefers brightness 30"). Do not reference any device or action outside the list above.
\end{tcolorbox}
\captionof{figure}{Prompt Stage 1}
\label{fig:prompt-stage-1}
\end{center}

\paragraph{Stage 2: Narrative Day-Plan Generation}


In the second stage, the generated persona memory cards are passed to a planning prompt for a discrete time window, as shown in \autoref{fig:prompt-stage-2}. The LLM generates a realistic natural-language narrative of the residents' activities, describing which devices they interact with, in what order, and the rationale behind their actions. Generating this intermediate narrative allows the model to effectively reason about physical locations and inter-resident coordination before producing any structured output, enabling the simulation to capture social behaviors, such as one resident keeping a room dim to avoid waking another.

\begin{center}
\centering
\begin{tcolorbox}[breakable, enhanced, sharp corners, colback=white, colframe=green!50, boxrule=1.0pt, left=1mm, right=1mm, top=1mm, bottom=1mm, nobeforeafter]
\small
Given the persona memory cards below, generate a realistic natural-language narrative of what \texttt{[resident names]} do between \texttt{[start time]} and \texttt{[end time]}. Describe which devices they interact with, why, and in which order. Account for who is at home and any relevant social constraints.

\medskip

CRITICAL CONSTRAINTS:\\
1. Exact Timestamps: Precede every specific device interaction in the narrative with an exact timestamp.\\
2. Explicit State Management: Whenever a resident leaves a room or finishes a task, explicitly state whether they turn the device off, change its state, or leave it running.\\
3. Definitive Language: Write definitively. Do not use words like "might", "may", or "if". State exactly what the residents do.\\
4. Specific Parameters: When a device state is changed (like brightness or temperature), state the exact value used based on the schema limits.\\
5. Schema Compliance: Use only the exact device names provided in the Stage 1 schema.

\medskip

Persona memory cards:
\texttt{[memory cards from Stage 1]}
\end{tcolorbox}
\captionof{figure}{Prompt Stage 2}
\label{fig:prompt-stage-2}
\end{center}

\paragraph{Stage 3: Action Extraction and Formatting}

In the third stage, the natural-language narrative generated in Stage 2 is passed to a strict formatting prompt, as shown in \autoref{fig:prompt-stage-3}. This prompt instructs the LLM to act as an action parser, extracting only the concrete device-level interactions and structuring them into a JSON array. Separating this formatting step from the actual day-plan generation reduces JSON syntax errors and prevents the model from hallucinating non-existent smart-home capabilities. To further improve robustness, Stage 3 is coupled with an API-level structured output specification that constrains the model to the target JSON schema. We additionally perform deterministic post-validation of required fields to verify output correctness.

\begin{center}
\centering
\begin{tcolorbox}[breakable, enhanced, sharp corners, colback=white, colframe=orange!60, boxrule=1.0pt, left=1mm, right=1mm, top=1mm, bottom=1mm, nobeforeafter]
\small
You are an action parser. Extract all smart-home device interactions from the narrative below into a JSON array.
\medskip

Each object must have the following fields: \\
- "timestamp" (string in HH:MM format)\\
- "resident" (string)\\
- "device" (string)\\
- "action" (string, exact match from schema)\\
- "action\_value" (integer, float, string, or null depending on the schema requirements)\\
- "intent" (string)
\medskip

Use only devices and actions from the following schema:\\
\texttt{[device schema JSON with separated actions and value formats]}
\medskip

CRITICAL CONSTRAINTS:\\
- Do not invent devices, actions, or parameters outside this schema.\\ 
- Separate the action from its parameter (e.g., "action": "set\_brightness", "action\_value": 45). If an action does not take a parameter (like "turn\_off"), set "action\_value" to null.\\
- Extract the exact timestamps provided in brackets in the narrative.
\medskip

Narrative:\\
\texttt{[Stage 2 output]}
\end{tcolorbox}
\captionof{figure}{Prompt Stage 3}
\label{fig:prompt-stage-3}
\end{center}

The resulting JSON array can be passed directly to an automated smart-home testbed to drive real device interactions and capture network traffic.

\paragraph{Stage 4: State and Memory Update}

To support multi-day or longitudinal simulations, the pipeline employs a rolling memory update after each time window. The LLM receives the JSON actions from Stage 3 alongside the existing persona memory cards and is asked to summarize any meaningful changes to residents' routines, preferences, or environmental state in a small number of bullet points per resident. These updated memory cards replace the previous ones and are fed back into Stage 2 as context for the next time window. This mechanism mirrors the reflective memory updates used in generative agents~\cite{park2023generative} and ensures spatial and behavioral consistency across extended simulations without exceeding LLM context windows.

\section{Proof of Concept}
As a proof of concept, we demonstrate the pipeline through a single illustrative run within one simulation window using OpenAI's gpt-5.4. This example is intended to highlight feasibility rather than robustness across repeated runs or configurations. Stage 4 remains part of the proposed architecture but is not instantiated in the subsequent evaluation.

\subsection{Experiment Setup}
Our experiment simulates two cooperative co-habitants, Alice, a work-from-home professional, and Bob, an office worker, in a compact urban apartment during a German winter morning. 
%

The device ecosystem consists of eight smart devices: Bedroom Lamp, Desk Lamp, Kitchen Lights, Vacuum Robot, Living Lamp, Smart Speaker, Thermostat, and Front Door Lock. We simulate the morning window from 06:00 to 10:00.

\begin{figure}[h]
    \centering
    \begin{tikzpicture}[
        node distance=1.0cm,
        font=\small\sffamily,
        data/.style={
            draw=black,
            thick,
            rounded corners=4pt,
            fill=white, 
            text width=8cm,
            inner sep=10pt,
            align=left
        },
        process/.style={
            ->,
            >=stealth,
            thick,
            draw=black!60
        },
        stageLabel/.style={
            draw=black,
            thick,
            rounded corners=3pt,
            font=\small\bfseries,
            inner sep=4pt
        }
    ]

    \node[data] (input) {
        \textbf{Input: Resident \& Device Schema} \\[1ex]
        \textbf{Bob:} A 9-to-5 Office Worker. He wakes up early at 06:00, needs to get ready quickly, eats breakfast, and departs the apartment for his commute by 08:30.\\[1ex]
        \texttt{\{"device": "Bedroom\_Lamp", "allowed\_actions": ["turn\_on", "turn\_off", "set\_brightness\_0\_to\_100", "set\_light\_temp\_2700K\_to\_6500K"]\}}
    };

    \node[data, below=of input] (memory) {
        \textbf{Stage 1 Output: Memory Card} \\[1ex]
        Likes a brighter, faster wake-up with \texttt{Bedroom\_Lamp}: \texttt{turn\_on}, \texttt{set\_brightness 60}, \texttt{set\_light\_temp 4200}.
    };

    \node[data, below=of memory] (narrative) {
        \textbf{Stage 2 Output: Narrative} \\[1ex]
        06:00 Bob turns on \texttt{Bedroom\_Lamp}, sets its brightness to 60, and sets its light temperature to 4200 so he can get dressed quickly.
    };

    \node[data, below=of narrative] (json) {
        \textbf{Stage 3 Output: Final JSON Schedule} \\[1ex]
        \texttt{[\{"timestamp": "06:00", "resident": "Bob", "device": "Bedroom\_Lamp", "action": "turn\_on", "action\_value": null, "intent": "get dressed quickly"\}, $\ldots$]}
    };

    \draw[process] (input) -- node[stageLabel, fill=blue!15] {Stage 1: Persona \& Context Init.} (memory);
    \draw[process] (memory) -- node[stageLabel, fill=green!15] {Stage 2: Narrative Day-Plan Gen.} (narrative);
    \draw[process] (narrative) -- node[stageLabel, fill=orange!15] {Stage 3: Action Extraction} (json);

    \end{tikzpicture}
    
    \caption{Example of the simulation pipeline processing a resident's morning routine.}
    \label{fig:pipeline_example}
\end{figure}
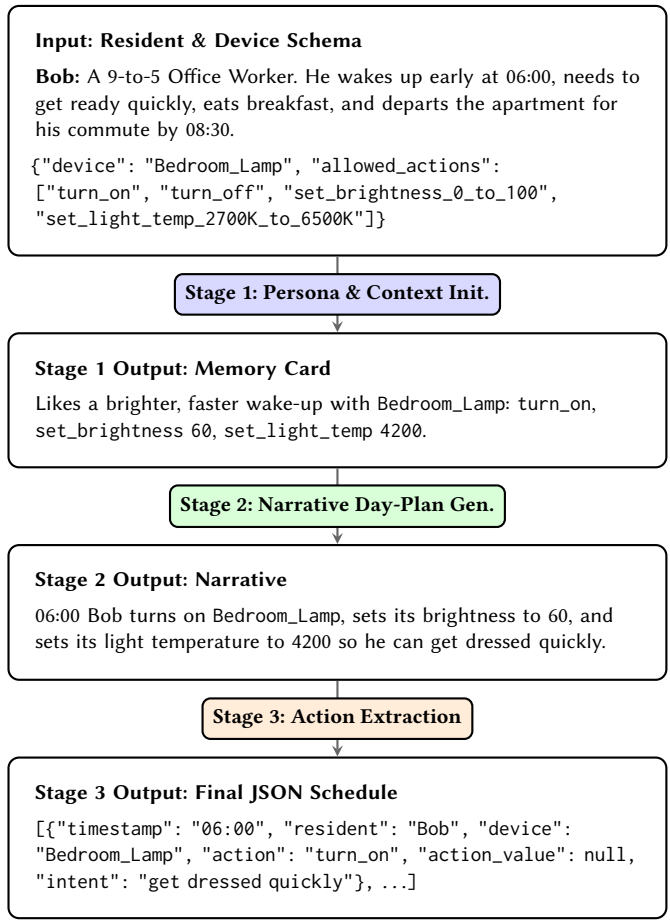

Figure~\ref{fig:pipeline_example} shows an exemplary excerpt from this experiment using Bob’s wake-up interaction with the Bedroom Lamp. 
The figure illustrates how one concrete moment from the simulation is processed through the pipeline, beginning with the resident description and device schema, continuing through the generated memory card and narrative, and ending in the structured JSON schedule used for execution.

\begin{figure}[htbp]
    \centering
    \includegraphics[width=0.99\columnwidth]{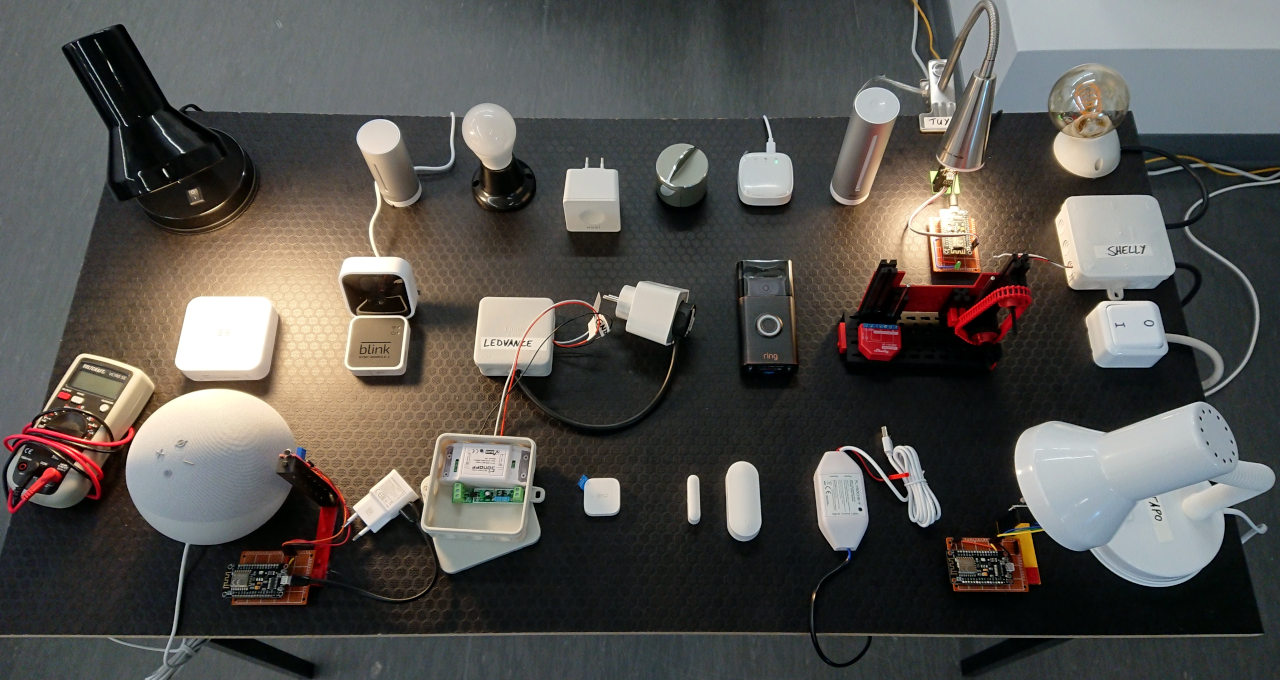}
    \caption{Our Smart-Home Testbed}
    \label{fig:testbed}
\end{figure}

Thus, our pipeline begins with Bob’s resident description together with the schema entry for the Bedroom Lamp. From this input, \textbf{Stage~1} generates a compact memory card that summarizes Bob’s relevant smart-home habit for waking up. \textbf{Stage~2} transforms this memory into a natural-language narrative event with a concrete timestamp. \textbf{Stage~3} then converts this narrative into a structured JSON action record that can be executed on our smart-home testbed, visible in Figure~\ref{fig:testbed}.  

\subsection{Observations}


In this minimal single-window instantiation, the refined pipeline produced contextually coherent device interactions grounded in both the residents’ respective routines and the environmental setting. Notably, the simulated outputs reflected the seasonal context: The narrative and extracted JSON captured Bob increasing the temperature setting of the \texttt{Smart\_Thermostat} upon waking up, and both residents relying heavily on indoor lighting (e.g., \texttt{Living\_Lamp} and \texttt{Desk\_Lamp}) due to the late winter sunrise.

Alice’s and Bob’s device interactions were also temporally distinct, with Alice’s usage spreading steadily across the morning and Bob’s actions being sharply clustered before his 08:30 departure. The generated JSON successfully matched the narrative and strictly respected the constraints of the provided device schema.








\section{Discussion}

In this work, we address the challenge of acquiring realistic smart-home data by simulating LLM-driven personas. Therefore, we introduce a novel approach to generate context-aware and human-centered device interaction schedules.

Our contribution is three-fold: First, we determine socio-technical dimensions to configure a smart-home household. Second, we propose and describe a simulation pipeline in detail. Third, we execute a proof-of-concept simulation, which demonstrates the ability of our pipeline to translate both high-level resident traits and environmental constraints into schema-compliant JSON schedules for physical testbeds. This indicates the possibility for LLMs to be utilized as a tool for generating behavioral smart-home data required for HCI, security, and privacy research.

\subsection{Practical and Design Implications}

We see three major implications for HCI research: 

\begin{itemize}
    \item \textbf{Direct testbed execution (Practical):} Because the pipeline outputs structured JSON schedules, the generated routines can be executed via automation systems on physical smart-home hardware. This allows security and privacy researchers to capture authentic network traffic and device states.
    \item \textbf{Standardized benchmarks (Practical):} The socio-technical dimensions facilitate the creation of off-the-shelf household configurations. If shared as open-source benchmarks, these synthetic personas could enable different research teams to evaluate their systems against baseline behaviors.
    \item \textbf{Rapid prototyping for multi-user dynamics (Design):} Simulating residents provides a prototyping tool for HCI practitioners. Designers can preemptively simulate 
    multi-user conflicts, like roommates with diverging morning routines or competing device intents to test how interfaces and automation logic handle these situations. 
\end{itemize}

\subsection{Limitations and Future Work}

While our proof-of-concept shows that the pipeline can generate temporally coherent and schema-compliant smart-home schedules, several limitations remain. Most importantly, we have not yet established the ecological validity of the generated routines, particularly whether LLM-generated personas can capture the irregular, inconsistent, or forgetful character of real domestic behavior. Moreover, our current evaluation is limited in scope and does not yet show how well the approach generalizes across diverse household configurations, device ecosystems, and longer simulation periods. Finally, although the pipeline is designed to produce execution-ready schedules, end-to-end execution on a physical smart-home testbed has not yet been demonstrated.

Future work will therefore focus on systematically evaluating the pipeline across a broader range of configurations, combining manual assessment with automated validation of output quality. We also plan to connect the pipeline to a live smart-home testbed to verify execution under real device constraints and, where ethically feasible, compare generated traces against real smart-home usage data. In the longer term, a primary direction for future work is to use this process as a foundation for building a large-scale, open-source dataset of simulated resident and device behaviors, while also supporting the generation of more realistic benign traffic for smart-home security research~\cite{juettner2026spartan}.




\section{Conclusion}
We presented a framework comprising five socio-technical dimensions and a multi-stage LLM pipeline for simulating smart-home residents together with their daily routines and translating those routines into executable device interaction schedules for physical testbeds. By simulating residents rather than traffic, the approach generates semantically grounded behavioral data, without requiring the surveillance of real households. An initial proof-of-concept demonstrated the feasibility of this approach, producing persona-consistent and schema-compliant interaction traces. As a work in progress, we see this as a first step toward a scalable, privacy-conscious simulation methodology that can serve the needs of HCI, security, and privacy research.

\section{AI Usage Disclosure}
To prepare this manuscript, AI tools were used for grammar correction and sentence editing.

\section{Acknowledgment}
 The authors acknowledge the financial support by the Federal Ministry of Research, Technology and Space of Germany and by “Sächsisches Staatsministerium für Wissenschaft, Kultur und Tourismus“ in the program Center of Excellence for AI-research “Center for Scalable Data Analytics and Artificial Intelligence Dresden/Leipzig“, project identification number: ScaDS.AI.


\bibliographystyle{ACM-Reference-Format}
\bibliography{literature}

\end{document}